\documentclass[letterpaper, 12pt]{article}

\usepackage{titlesec}% http://ctan.org/pkg/titlesec
\usepackage{graphicx}
\usepackage[version=3]{mhchem} % Formula subscripts using \ce{}
\usepackage{graphicx}
\usepackage{subcaption}
\usepackage{array,multirow}
\usepackage{amsmath}
\usepackage{todonotes}
\usepackage{algpseudocode,algorithm,algorithmicx}

\usepackage{xcolor}

\usepackage{cmjStyle} %use CMJ style
\usepackage{natbib} %natbib package, necessary for customized cmj BibTeX style
\bibpunct{(}{)}{;}{a}{}{,} %adapt style of references in text
\raggedright % use this to remove spacing and hyphenation oddities
\begin{document}

{\cmjTitle A framework to compare music generative models using automatic evaluation metrics extended to rhythm.}
\vspace*{24pt}

{\cmjAuthor Sebastian Garcia-Valencia\textsuperscript{a,b}, Alejandro Betancourt\textsuperscript{c} and Juan G. Lalinde-Pulido\textsuperscript{a}}
\newline
\newline
\textsuperscript{a}Computer Science Department, Universidad EAFIT, Medellin, Colombia
\newline
\textsuperscript{b}Research and Development, AIVA Technologies, Luxembourg City, Luxembourg
\newline
\textsuperscript{c}Digital Department, Ecopetrol, Bogota, Colombia

\begin{abstract}
To train a machine learning model is necessary to take numerous decisions about many options for each process involved, in the field of sequence generation and more specifically of music composition, the nature of the problem helps to narrow the options but at the same time, some other options appear for specific challenges. This paper takes the framework proposed in a previous research that did not consider rhythm to make a series of design decisions, then, rhythm support is added to evaluate the performance of two RNN memory cells in the creation of monophonic music. The model considers the handling of music transposition and the framework evaluates the quality of the generated pieces using automatic quantitative metrics based on geometry which have rhythm support added as well.
\end{abstract}

\section{keywords}
Automatic Evaluation; Rhythm; Framework; Monophonic Music; Generative Model; Recurrent Neural Network

\section{Introduction} \label{sec:introduction}

Artificial intelligence is a term appearing everywhere in the last years, the variety of fields it is transforming is huge and the examples of breaking advances in everything it is applied to seems not to stop. Music composition is not the exception, every day the level of compositions generated by AI gets only better, and more and more composers begin to integrate AI-powered systems in their arsenal of tools.

With the development of the field, comes the fact that there are too many options for each aspect of the development of a machine learning model, i.e. datasets, representations, objective functions, hyperparameters, etc. Therefore the importance of a way to formally narrow these options.

We can characterize the challenges for this research in two categories. On one hand, we have the algorithmic challenges that any machine learning problem has (which data to use, which algorithmic architecture, how to train the model). On the other hand, we must consider some further challenges specific to the field of music composition (how to represent the music, how to keep long term consistency, how to deal with transposition, and how to evaluate the quality of the compositions)

In previous research done in \citep{garciavalencia2020sequence}, we proposed a framework to progressively compare and justify a good set of combinations for these options in the case of a sequence generator of melodies without rhythm (all notes are quarter notes). Here we will apply the same framework adding support for rhythm as follows, getting a whole monophonic music generator.

\textbf{Training Data: } Music datasets are usually the result of scraping music repositories. The most popular container in recent years is the MuseScore sheet music archive \citep{MuseScore2002}, because there is a significant community supporting and contributing to it, an example of a dataset of monophonic music with pieces of musescore is the mono-musicxml-dataset \citep{DBLP:journals/corr/WelU17}. We use an updated version of the mono-midi-transposition-dataset already used in the previous research but with support for rhythm.

\textbf{Algorithmic Architecture: } 

Just like in the previous research, we use an architecture based on Recurrent Neural Networks with memory mechanisms but defined and implemented from scratch. It has a multicell RNN layer where n units of a certain type of RNN cell can be set, the election of this number and type is not trivial.

\textbf{Objective Function: } Concerning the optimization of the model in training, usually, cross-entropy is a popular choice for classification and generative models \citep{Johnson2017, Hutchings2017, whorley2016, kawthekarevaluating}. The behavior of this metric is used as criteria to choose the number of units in the multicell RNN since it was already tested that optimization of cross-entropy corresponds with improvements in the music generation \citep{garciavalencia2020cross}.

Respect to the musical aspects:

\begin{enumerate}
    \item{ \textbf{Music Representation:} The notes are transformed into tuples (p,d), where p is the pitch as midi value and d is the duration, then they are encoded as embeddings to add semantic meaning as tested in \citep{garciavalencia2020embeddings}.
}
    \item{ \textbf{Long Term Consistency:} A common problem in sequence generation is that, after some iterations, the generated sequence loses sense and becomes random data. To address this problem we analyze the performance of two different types of memory cells:
    \begin{enumerate}
    \item LSTM \citep{Hochreiter1997}: The Long-Short-Term Memory NN is composed of 4 elements: a cell state which determines the current context, a forget gate which decides which information remove, an input gate to insert new information, and an output gate to decide the output \citep{Olah2015f}.  
    \item GRU \citep{2014arXiv1406.1078C}: The Gated Recurrent Unit is very similar to the LSTM. It has two gates, the update gate which is a combination of the forget and input gates in the LSTM, and the reset gate which decides the rate of past information kept. Training a GRU network is usually faster because it has fewer tensors than LSTM.
\end{enumerate}
The authors in \cite{DBLP:journals/corr/ChungGCB14} compare LSTM and GRU and conclude that GRU performs better in 3 out of 4 datasets, however, there is no consensus about which is better and it is normal to try both to find which fits better the use case.}
    \item{ \textbf{Music Transposition:} Music transposition makes two pieces of music to be perceived in essence the same song despite having almost all notes different. The updated version of the dataset with time support has the same 3 strategies variations to address music transposition: i) A control case where the sequences are the notes, ii) A modified case storing the pitch intervals iii) The last case on which each sequence is augmented to 12.}
    \item{ \textbf{Music Evaluation:} According to the recent bibliography, music evaluation can be divided into two groups. The first one uses mathematical formulations \citep{Colombo2017, Jaques2016, Tymoczko2011}, while the second one uses musical experiments to consider human judgments about the quality of the composition.
    To quantify the quality of the compositions, we add rhythm support to the three quantitative metrics proposed in the previous research (Conjunct Melodic Motion, Limited Macroharmony and Centricity) which considers indications of tonality based on music theory and geometry. The proposed metrics are used to compare different combinations of hyperparameters and the impact of the Music Transposition strategies.  See section "\nameref{subsec:quantmetricdescr}" for a further description of the metrics.

}
\end{enumerate}

\textbf{The Framework: }In summary, coming from the conclusions of the previous research \citep{garciavalencia2020sequence} we already chose the mono-midi-transposition-dataset for training purposes, the general architecture based in a multi-cell RNN, cross-entropy as objective function, tuples as music representation, embeddings as encoding and quantitative metrics for quality evaluation. we still have 3 options of dataset strategy for transposition, 2 types of memory cells and n numbers of units to put in the multicell RNN for long-term consistency.
First, we will use the cross-entropy convergence of the models in training and validation to choose the number of units, and then the quantitative metrics to evaluate the datasets and memory cell types.

The remaining part of this paper is organised as follows: Chapter "\nameref{sec:generatingMelodies}", introduces our approach for the RNN architecture, the training phase, and the music evaluation. Chapter "\nameref{sec:experiments}", presents the experiments to understand the selected architectures and finally tune the hyperparameters of the proposed networks (e.g., Number of units, Memory Cell). Chapter "\nameref{sec:results}", analyses the models and the generated melodies from the quantitative and musical perspective. Finally, chapter "\nameref{sec:conlusions}", concludes and provides some future research lines of this work.

\section{Generating melodies} \label{sec:generatingMelodies}

Following the ideas from section "\nameref{sec:introduction}", our goal is to develop a network to generate melodies. Section "\nameref{subsec:rnnarchitecture}", describes the architecture and training of the RNN. Section "\nameref{subsec:dataset}" and "\nameref{subsec:quantmetricdescr}", illustrates respectively the updates to the dataset and quantitative metrics respect to the previous research. The result of applying these metrics to the dataset is used as the baseline for the performance analysis of section "\nameref{subsec:automEval100songs}".
\newline

\subsection{RNN Architecture} \label{subsec:rnnarchitecture}

This model with time support was done from scratch and do not reuse the one of the previous paper, however, its high-level logic is very similar.

\begin{figure}[h!]
    \centering
    \includegraphics[width=\textwidth]{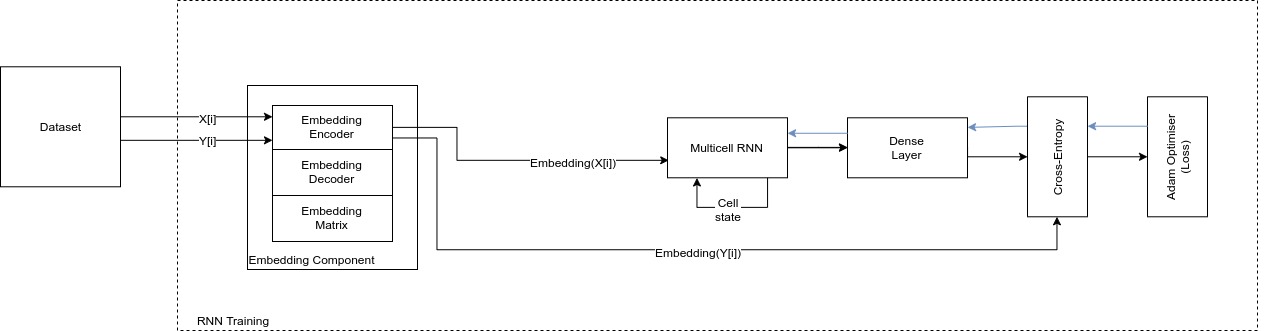}
    \caption{RNN Architecture training mode}
    \label{fig:rnnarchitecturetraining}
\end{figure}

\begin{figure}[h!]
    \centering
    \includegraphics[width=\textwidth]{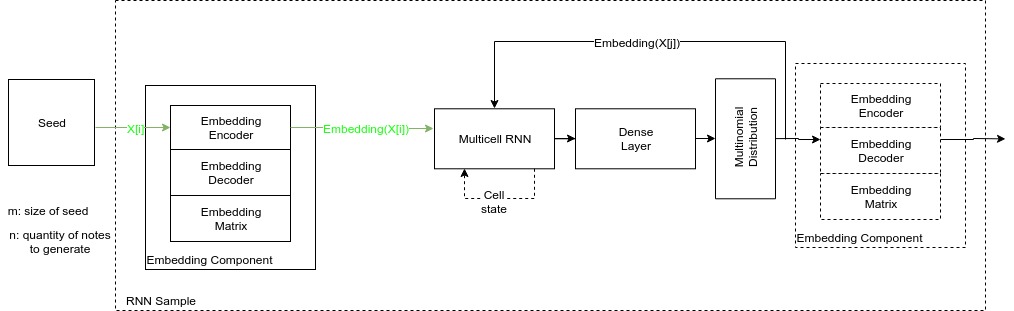}
    \caption{RNN Architecture sampling mode}
    \label{fig:rnnarchitecturesampling}
\end{figure}

The workflow through the network varies depending on the use, namely training mode (fig. \ref{fig:rnnarchitecturetraining}) or sampling mode (fig. \ref{fig:rnnarchitecturesampling}). For training, the starting point is the dataset. First, every sample goes to the embedding component to be encoded as a vector.

The embedding goes to the multicell RNN, which can contain an array of cells of any type (LSTM or GRU). Once the embedding flows through the multicell RNN, the internal state of the cells changes and goes as feedback input for the next iteration. After that, there is a densely connected layer. 

Using the output of the dense layer and the encoded Y, the network calculates the cross-entropy. Using an Adam optimiser, this last layer back-propagates the error through all the layers (light lines in fig. \ref{fig:rnnarchitecturetraining}).

In the case of sampling mode, the starting point is the seed, which is the sequence that the network will use to initiate its internal state and then generate new notes. From here, the sampling mode has two phases. In the first one (light lines in fig. \ref{fig:rnnarchitecturesampling}), each sample of the seed flows through the embedding encoder to become its vector representation and then change the internal state of the cells in multicell RNN. Once it finishes with all the samples in the seed, phase 2 begins (dark lines in fig. \ref{fig:rnnarchitecturesampling}). The output of the multicell flows through the dense layer and a multinomial distribution layer which outputs the most probable next embedding. This output goes back as an input to the multicell RNN in the next iteration, and simultaneously it is decoded and added to the generated melody. This process repeats n times.

\subsection{Mono-midi-transposition-dataset} \label{subsec:dataset}

The dataset\footnote{The general dataset is available in \url{https://sebasgverde.github.io/mono-midi-transposition-dataset/}} used in this paper is an updated version of the dataset used in the previous research. The process consists of the same 3 steps\footnote{All the code, files and final datasets to replicate the results of the whole paper are available in \url{https://sebasgverde.github.io/rnn-time-music-paper/}} than the previous paper ( i) Scraping,  ii) Preprossessing and iii) Cleaning) with the difference that in the preprocessing, instead of just an array containing the sequence of notes, each midi is transformed into an array containing a list of tuples (p,d), where p is the pitch as midi value and d is the duration using a bar resolution of 16, this means that a whole note is represented with a value of 16 and a 16th note with a value of 1. Also, in the cleaning, only pieces with at least 12 notes and with durations of max 16 are kept. 

The final list of arrays is used as the base to create 3 datasets (fig \ref{fig:dataset}.4 and \ref{fig:dataset}.5): 

\begin{figure}[h!]
    \centering
    \includegraphics[width=\textwidth]{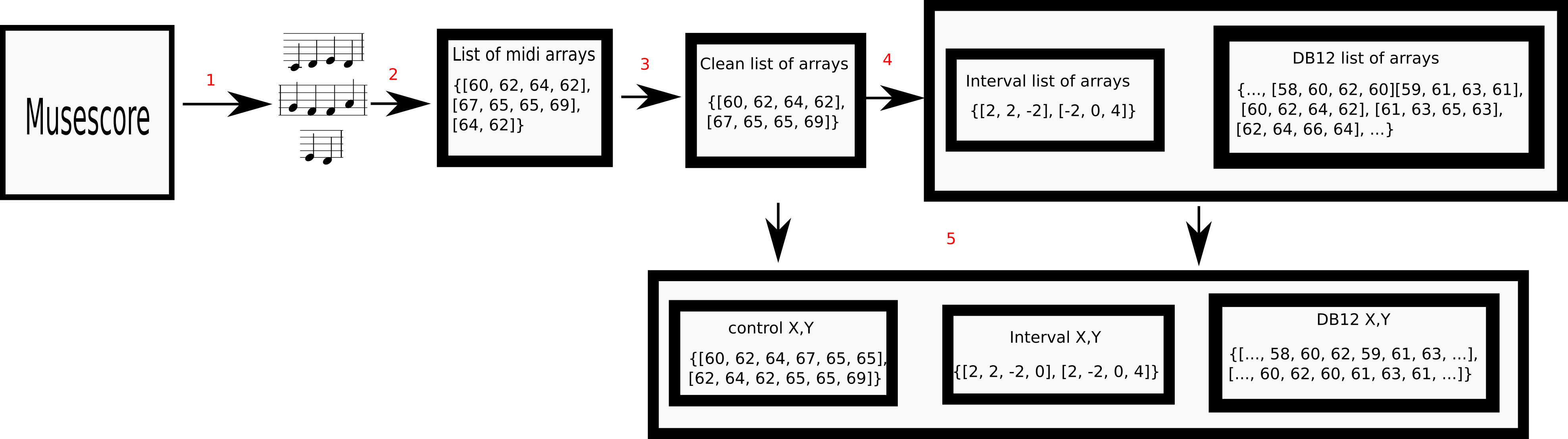}
    \caption{Dataset workflow}
    \label{fig:dataset}
\end{figure}

\textbf{Control Dataset: }This is the base case, on which X is the concatenated array of the original songs and Y is X shifted by one position.

\textbf{DB12 Dataset: }To construct this dataset, each song is transposed 12 times (on each degree of the chromatic scale).

\textbf{Intervals Dataset: }In this case, we do not have a sequence of notes, but a sequence of relative changes. 

\subsection{Description of the Quantitative Evaluation} \label{subsec:quantmetricdescr}

Here we define the adaption of the 3 metrics (i.e. Conjunct Melodic Motion, Limited Macroharmony and Centricity) proposed in \cite*{Tymoczko2011} and implemented in the previous research \citep{garciavalencia2020sequence} to evaluate automatically the quality of the generated pieces measuring the tonality.  Since we need to support rhythm, the calculation of the span changes respect to the previous research.

Let's convert the melody into an array of resolution 16 (16 elements per bar), where the pitch is in the position of the onsets and all other elements are zero. Let's define the $span_{i}$ as a subset of song from $X_{i}$ to $X_{i+n}$, where n = $|$span$|$ is the size of the span (in this case the constant 32, the size of two bars), m is the size of the step we use to move through the array (in this case 4, the size of a quarter note) and  $Sq$ the quantity of spans given by equation \ref{eq:equationquantityspans}.

\begin{equation}
    Sq= 
\begin{cases}
    1, & \text{if } |song| \leq n\\
    \frac{|song|-n}{m}+1, & \text{otherwise}
\end{cases}
\label{eq:equationquantityspans}
\end{equation}

\begin{enumerate}
    \item{ \textbf{Conjunct Melodic Motion (CMM):} CMM looks for smooth transitions through the melody, in other words, the changes between notes must have short intervals.}
    \item{ \textbf{Limited Macroharmony (LM):} 
    Macroharmony measures the diversity of notes in melodies. This is captured by measuring the number of different notes in short slices of time (spans).}
    \item{ \textbf{Centricity (CENTR):} Centricity claims that in short slices of time (spans), there must be a note which appears with more frequency than the others.}
\end{enumerate}

\subsubsection{Geometry descriptive statistics for dataset}

We apply now these metrics to the database to have a baseline to compare. As table \ref{tab:metricDataset} shows, for the CMM and the LM the average is above the perfect score of 1 by 1.23 and 1.05 respectively, this evidence that a score above 1 in the generated pieces is not necessarily a bad result. In the case of centricity, we can conclude that a tonal melody should have a note that is at least the 27\% of the melody.

\begin{table}[h!]
\centering
\begin{tabular}{ccc} \hline
CMM     & LM      & CENTR \\ \hline
% mean dev std     & mean dev std      & mean dev std  \\         
2.23 $\pm$ 0.98 & 2.05 $\pm$ 1.18 & 0.27 $\pm$ 0.14 \\
\hline
\end{tabular}

\caption{Dataset evaluation}\label{tab:metricDataset}
\end{table}

\section{Experiments} \label{sec:experiments}

This section compares different types of RNN cells (2), with different quantities of units in the multicell RNN layer (5) and different transposition strategies (3). The combination of these decisions gives 30 different models to be analysed. This section defines the optimal number of units, for every data-set and memory cell, reducing the number of models to 6 (3 datasets and 2 memory cells). Looking for simplicity, the experiments are grouped by data-set, we try powers of two number of units beginning with 128 (i.e. 128, 256, 512, 1024, 2048). The convergence of the cross-entropy is used as evaluation metric looking for a balance between a good training and validation loss value. 

Regarding the training conditions: i) In all the cases the batch size is 64 and the sequence length is 100, which means that every training iteration uses 64 sequences of 100 notes simultaneously. ii) For the control and the interval dataset, the model trains 200 epochs, which means that uses the complete dataset 200 times. For the DB12 dataset, which is 12 times bigger than the other 2, it trains only 90 epochs. 

\subsection{Control dataset experiment}

\begin{figure}[h!]
    \centering
    \begin{subfigure}[b]{0.4\textwidth}
        \includegraphics[width=\textwidth]{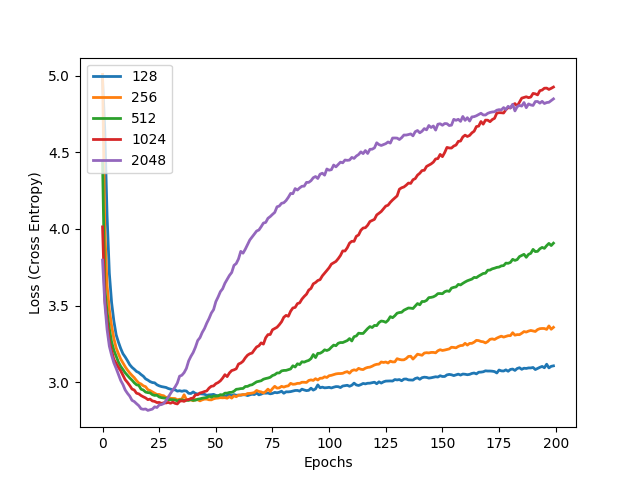}
        \caption{LSTM validation}
        \label{fig:controlvallstm}
    \end{subfigure} 
    \begin{subfigure}[b]{0.4\textwidth}
        \includegraphics[width=\textwidth]{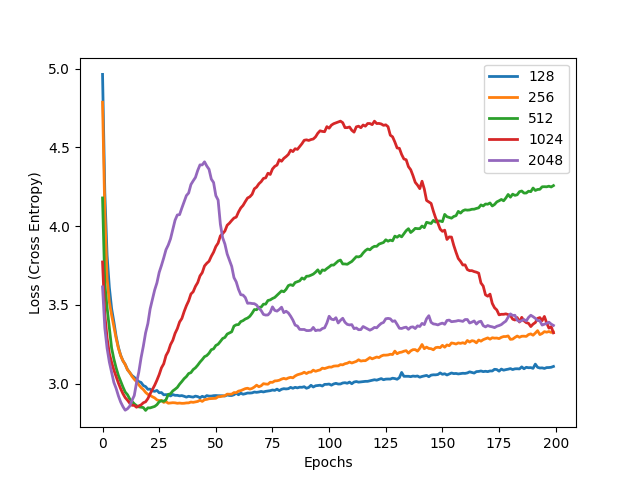}
        \caption{GRU validation}
        \label{fig:controlvalgru}
    \end{subfigure}
    
    \begin{subfigure}[b]{0.4\textwidth}
        \includegraphics[width=\textwidth]{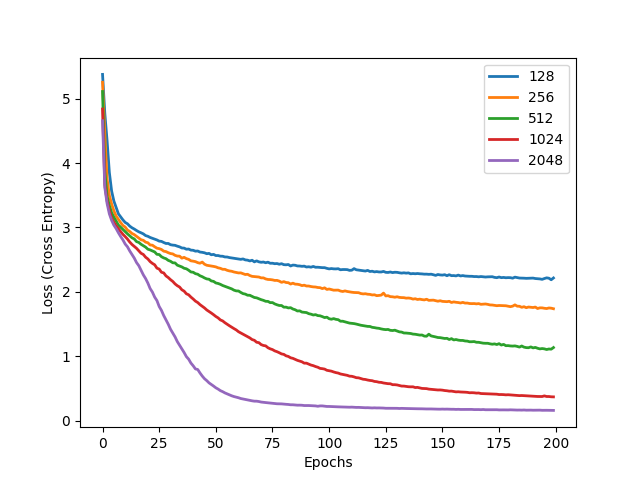}
        \caption{LSTM training}
        \label{fig:controltrainlstm}
    \end{subfigure}     
    \begin{subfigure}[b]{0.4\textwidth}
        \includegraphics[width=\textwidth]{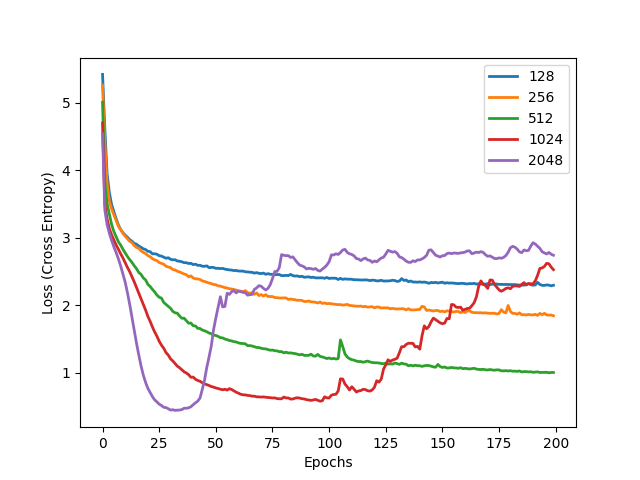}
        \caption{GRU training}
        \label{fig:controltraingru}
    \end{subfigure}

    \caption{Control Dataset Experiment Learning Curves}\label{fig:control}
\end{figure}

The optimal number of units for the LSTM model is 2048, after epoch 60 the training loss stops improving (Fig. \ref{fig:controltrainlstm}) while the validation loss continues getting worse (Fig. \ref{fig:controlvallstm}), this is a clear insight of overfitting, however, in the context of sequence generation, overfitting doesn't have the same meaning that it has in other tasks like classification, and therefore even if is a good criteria to choose a model and train step, it is not the definitive one. 

In the case of LSTM, the 2048 model has a stable behavior and the minimal training loss, we chose this model around epoch 60, where the validation loss is minimal.

Even if the GRU model with 2048 units has the minimal training loss, it is too unstable (Fig. \ref{fig:controltraingru}), therefore we use the 1024 units one around epoch 50 which is more stable and shows less overfitting (Fig. \ref{fig:controlvalgru})

\subsection{Interval dataset Experiment}

\begin{figure}[h!]
    \centering
    \begin{subfigure}[b]{0.4\textwidth}
        \includegraphics[width=\textwidth]{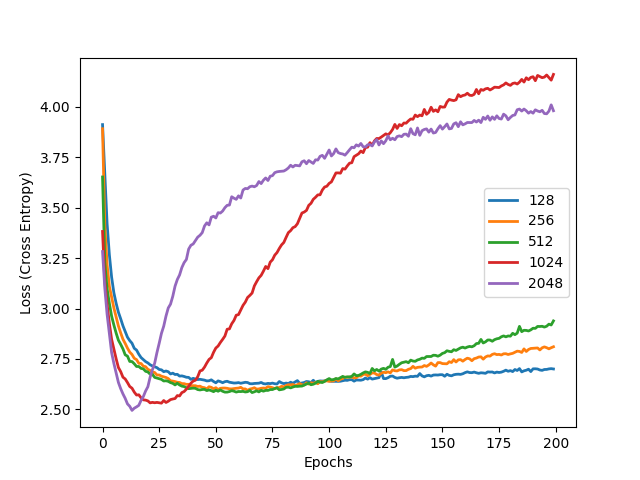}
        \caption{LSTM validation}
        \label{fig:intervalvallstm}
    \end{subfigure} 
    \begin{subfigure}[b]{0.4\textwidth}
        \includegraphics[width=\textwidth]{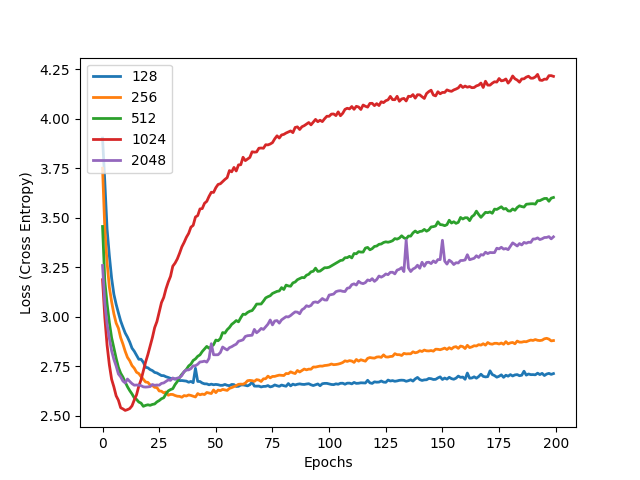}
        \caption{GRU validation}
        \label{fig:intervalvalgru}
    \end{subfigure}
    
    \begin{subfigure}[b]{0.4\textwidth}
        \includegraphics[width=\textwidth]{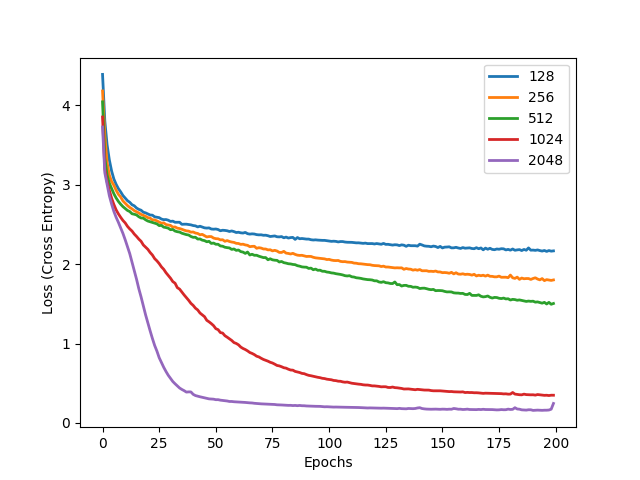}
        \caption{LSTM training}
        \label{fig:intervaltrainlstm}
    \end{subfigure}     
    \begin{subfigure}[b]{0.4\textwidth}
        \includegraphics[width=\textwidth]{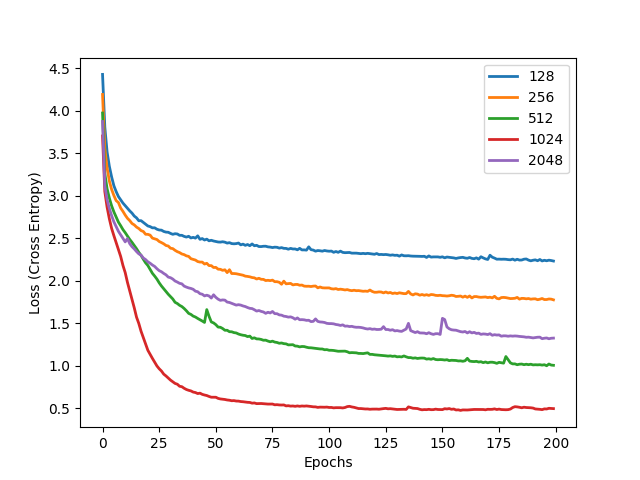}
        \caption{GRU training}
        \label{fig:intervaltraingru}
    \end{subfigure}    
    
    %   Number of layers: [\textcolor{orange}{\textbullet}] 1 [\textcolor{red}{\textbullet}] 2 [\textcolor{magenta}{\textbullet}] 3 [\textcolor{blue}{\textbullet}] 4 [\textcolor{cyan}{\textbullet}] 5
    \caption{Interval dataset Experiment Learning Curves}\label{fig:interval}
\end{figure}

For the interval dataset case we use the same reasoning for both memory cells, the minimal training loss in a train step of minimal validation loss. In the case of LSTM is the 2048 units model around epoch 60, even if the 1042 units one is close (Fig. \ref{fig:intervaltrainlstm}) the validation loss shows a worse behavior (Fig. \ref{fig:intervalvallstm}).

The behavior of the GRU model for this dataset shows interesting results, the 2048 units model never reaches the minimum value as it did with the control dataset where it did it at least for a while before becoming unstable, in this case, 512 and 1024 are better for the training (Fig. \ref{fig:intervaltraingru}), we can notice a tendency of the GRU cells to have some unexpected behaviors.

\subsection{DB12 Dataset Experiment}

\begin{figure}[h!]
    \centering
    \begin{subfigure}[b]{0.4\textwidth}
        \includegraphics[width=\textwidth]{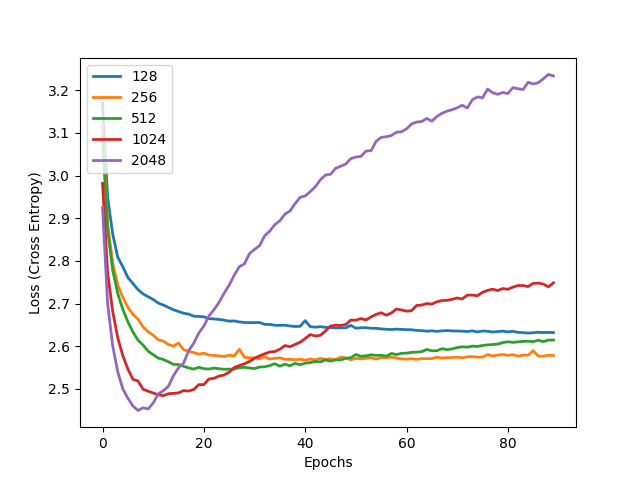}
        \caption{LSTM validation}
        \label{fig:db12vallstm}
    \end{subfigure} 
    \begin{subfigure}[b]{0.4\textwidth}
        \includegraphics[width=\textwidth]{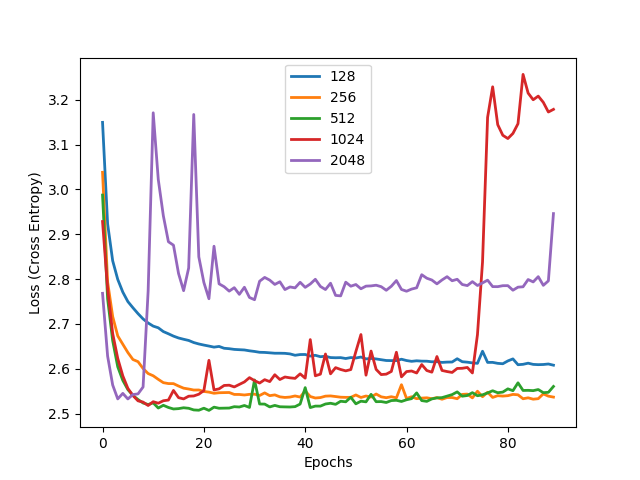}
        \caption{GRU validation}
        \label{fig:db12valgru}
    \end{subfigure}
    
    \begin{subfigure}[b]{0.4\textwidth}
        \includegraphics[width=\textwidth]{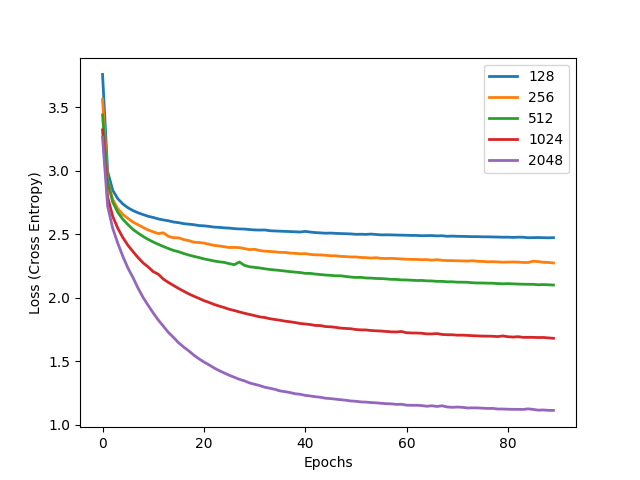}
        \caption{LSTM training}
        \label{fig:db12trainlstm}
    \end{subfigure}     
    \begin{subfigure}[b]{0.4\textwidth}
        \includegraphics[width=\textwidth]{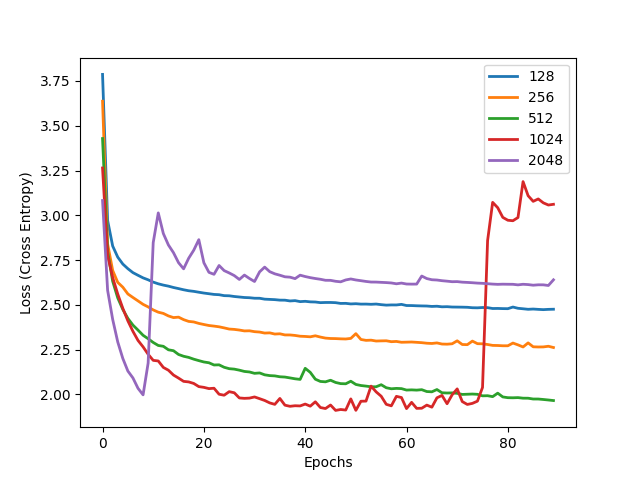}
        \caption{GRU training}
        \label{fig:db12traingru}
    \end{subfigure}

    \caption{DB12 Dataset Experiment Learning Curves}\label{fig:db12}
\end{figure}

In the LSTM experiment (Fig. \ref{fig:db12trainlstm}), results are similar to the other two datasets, with correspondence between training loss and units count, the 2048 unit model around epoch 50 is the chosen in this case.

GRU DB12 model has a bad performance in this experiment as well, showing a general difference in stability for dense models between LSTM and GRU. The chosen model, in this case, is the 512 units one in the minimal training loss epoch, this is because it is very close to the 1024 units model (Fig \ref{fig:db12traingru}) but shows more stability and less overfitting in the validation loss (Fig. \ref{fig:db12valgru}).

\subsection{Analysing the Network Depths}

Table \ref{tab:optimalnumberoflayers} summarises the optimal number of units for each dataset and memory cell.  The following part of this section analyses the advantages and disadvantages of each dataset (rows) and the performance of each memory-mechanism (columns). The remaining part of this paper uses the number of units defined in Table \ref{tab:optimalnumberoflayers}.

\begin{table}[h!]
\centering
\begin{tabular}{lccc} \hline
         & LSTM     & GRU \\ \hline
CONTROL  & 2048 & 1024 \\
INTERVAL & 2048 & 1024 \\
DB12     & 2048 & 512 \\
\hline
\end{tabular}

\caption{Optimal number of units per model}\label{tab:optimalnumberoflayers}
\end{table}

Regarding the datasets, it is interesting, that the \textit{DB12 dataset} is the only one with 512 units as the best option, it is as well, the only dataset where none of the models reach a training loss below 1 within the first 50 epochs, this suggests that models take longer to converge with this dataset, but this can be explained by the fact that the DB12 dataset is 12 times bigger than the other two.

Regarding the type of memory cell, the LSTM shows a very stable behavior across the datasets, always having a direct correspondence between unit count and loss convergence, as well as a general stable behavior of the curves. The GRU model, on the other hand, shows instability in the curves, especially for dense models like the ones with 1024 and 2048 units.

\section{Results} \label{sec:results}

This chapter introduces the evaluation and analysis of the 6 models of table \ref{tab:optimalnumberoflayers}. In section "\nameref{subsec:automEval100songs}", the 3 metrics described in "\nameref{subsec:quantmetricdescr}", test 100 songs for each of the 6 models to quantify their composition capabilities. Section "\nameref{subsec:generatedmelodies}", visualises the 6 most representative songs, and provide analysis from the musical viewpoint.

\subsection{Analysis of Metrics in the Models}\label{subsec:automEval100songs}

To evaluate the 6 models from table \ref{tab:optimalnumberoflayers}, each model generates 100 songs adding 30 new notes to a seed of a D4 half note followed by an E4 quarter note and an F4 quarter note. Table \ref{tab:metricsSelectedModel} shows the average and standard deviation for each metric and model.

\begin{table}[h!]
\centering
\begin{tabular}{clccc} \hline
      &  & LSTM     & GRU    \\ \hline
\multirow{ 2}{*}{CMM} &  
control & 2.06 $\pm$ 0.66 & 2.01 $\pm$ 0.53 \\
&interval  & \textbf{1.65 $\pm$ 0.61} & 2.21 $\pm$ 0.73 \\
&db12  & 2.30 $\pm$ 0.89 & 1.66 $\pm$ 0.65 \\
\hline
\multirow{ 3}{*}{LM} &  
control & 1.97 $\pm$ 0.60 & 1.84 $\pm$ 0.55 \\
&interval  & \textbf{1.67 $\pm$ 0.70} & 1.99 $\pm$ 0.70 \\
&db12  & 2.25 $\pm$ 1.02 & 2.20 $\pm$ 0.87 \\
\hline
\multirow{ 3}{*}{CENTR} &  
control & 0.30 $\pm$ 0.10 & 0.28 $\pm$ 0.10 \\
&interval  & 0.24 $\pm$ 0.11 & 0.27 $\pm$ 0.11 \\
&db12  & 0.30 $\pm$ 0.14 & \textbf{0.34 $\pm$ 0.17} \\
\hline

\end{tabular}

\caption{Means and standard deviation for the 100 songs generated by the final models with the 3 metrics.}\label{tab:metricsSelectedModel}
\end{table}

According to the Conjunct Melodic Motion (CMM) the closest model to 1 is the \textit{INTERVAL-LSTM} (1.65) outperforming the average of the dataset (2.23), close to it is the \textit{DB12-GRU}(1.66) but with a higher standard deviation. The \textit{DB12-LSTM} model shows a significantly worse result, and the largest standard deviation, positioning it as the least reliable model according to the CMM.

In the case of LM the DB12 dataset has the worse performance in general for this metric with both values above the dataset average (2.05). The model with the best LM is the \textit{INTERVAL-LSTM} with 1.67, which also outperforms the 2.05 of the dataset (table \ref{tab:metricDataset}). 

Regarding the centricity, The model which shows higher centricity is the \textit{DB12-GRU}. Notice that this model has poor LM, this makes sense, while centricity measures the prevalence of a note, the LM, makes the opposite. The bad performance of this model in the LM metric indicates a strong repetition of notes

In general, INTERVAL models show a low centricity (tends to use more notes), which is evident in the good results for LM. Between the 2 strategies for transposition learning, the interval representation is the most stable.

With respect to the best model for the 3 metrics, the \textit{CONTROL-GRU} model shows the best trade-off for the three metrics. It is the only model that outperforms the centricity of the dataset without affecting the LM considerably. In summary, concerning only the quantitative analysis, the best models are i) \textit{INTERVAL-LSTM} for CMM and LM, ii) \textit{DB12-GRU} for CENTR and iii) \textit{CONTROL-GRU} for general tonality.

\subsection{Generated Melodies Musical Analysis} \label{subsec:generatedmelodies}

Using the average metrics for the 100 songs generated by each of the 6 models (Table \ref{tab:metricsSelectedModel}), it is possible to select a representative song per model using the Euclidean Distance to the average as selection criteria. Table \ref{tab:metricsMostRepresSongs} summarises the metrics for the selected songs.

\newcommand{\rotationtalbeangle}{60}
\setlength{\tabcolsep}{3pt}
\begin{table}[h]
\centering
  \begin{tabular}{lcccccc}
\cline{2-7}
 & \multicolumn{3}{c}{LSTM} & \multicolumn{3}{c}{GRU} \\\cline{2-7}           %% <--  Changed
 & \rotatebox[origin=c]{\rotationtalbeangle}{CMM}& \rotatebox[origin=c]{\rotationtalbeangle}{LM}& \rotatebox[origin=c]{\rotationtalbeangle}{CENTR}    
 & \rotatebox[origin=c]{\rotationtalbeangle}{CMM}& \rotatebox[origin=c]{\rotationtalbeangle}{LM}& \rotatebox[origin=c]{\rotationtalbeangle}{CENTR} \\
 \hline
CONTROL  & 1.94 & 2.00 & 0.30 & 2.00 & 1.80 & 0.26 \\
INTERVAL  & 1.62 & 1.67 & 0.32 & 2.16 & 2.07 & 0.24 \\
DB12  & 2.22 & 2.33 & 0.36 & 1.62 & 2.14 & 0.29 \\

\hline
\hline
\end{tabular}

\caption{The 3 metrics for the most representative of the 100 songs generated by the final models in each case.}\label{tab:metricsMostRepresSongs}
\end{table}

\begin{figure}[h]
    \centering
    \begin{subfigure}[b]{\textwidth}
        \includegraphics[trim= 0 2000 0 200,clip,width=\textwidth]{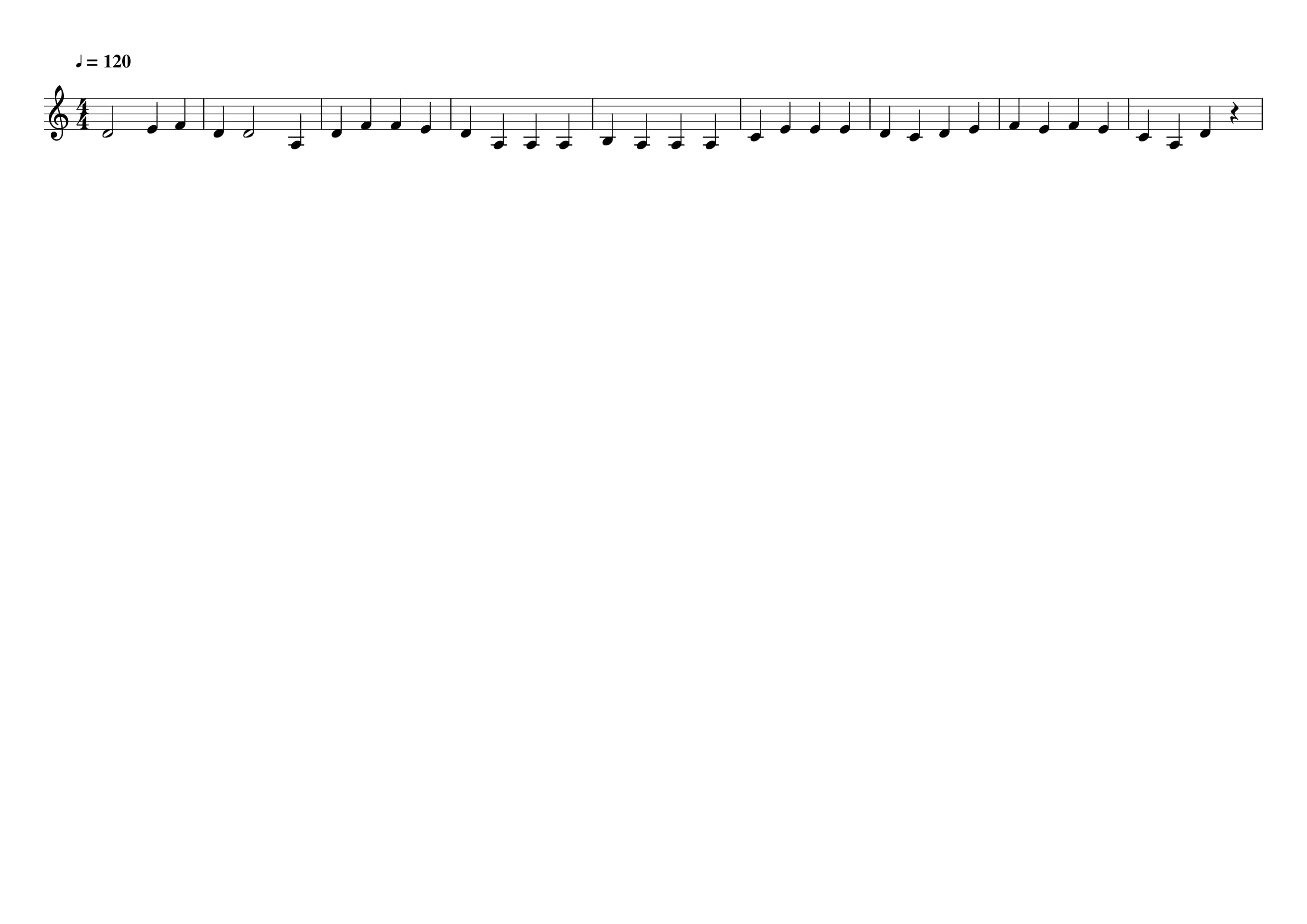}
        \caption{LSTM Model}
        \label{fig:baselstmsong}
    \end{subfigure}
    
    ~ %add desired spacing between images, e. g. ~, \quad, \qquad, \hfill etc. 
    \begin{subfigure}[b]{\textwidth}
        \includegraphics[trim= 0 2000 0 200,clip,width=\textwidth]{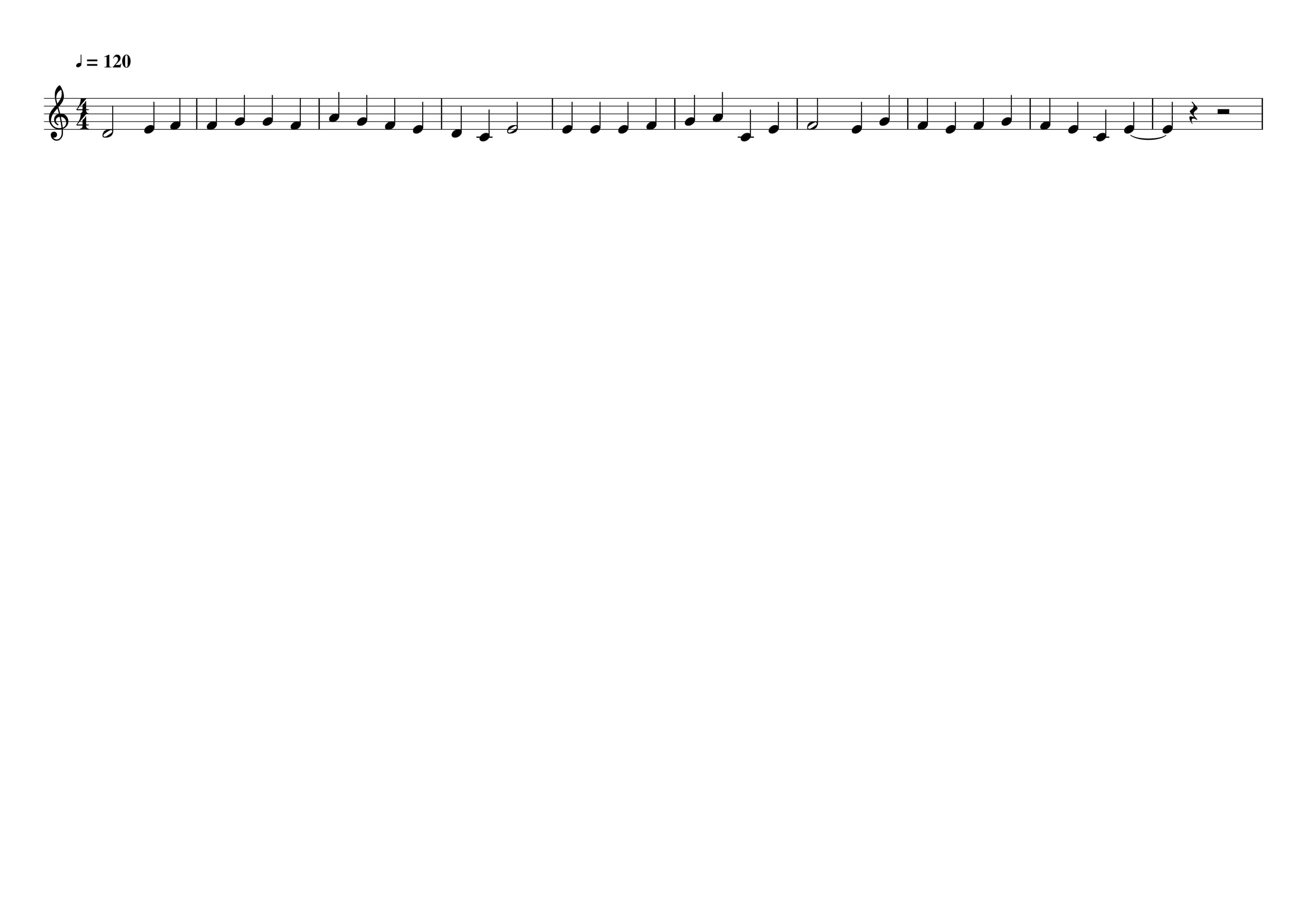}
        \caption{GRU Model}
        \label{fig:basegrusong}
    \end{subfigure} \\
    \caption{Generated Melodies Control Dataset}\label{fig:generatedmelodiescontrol}
\end{figure}

\begin{figure}[h]
    \centering    
    \begin{subfigure}[b]{\textwidth}
        \includegraphics[trim= 0 2000 0 200,clip,width=\textwidth]{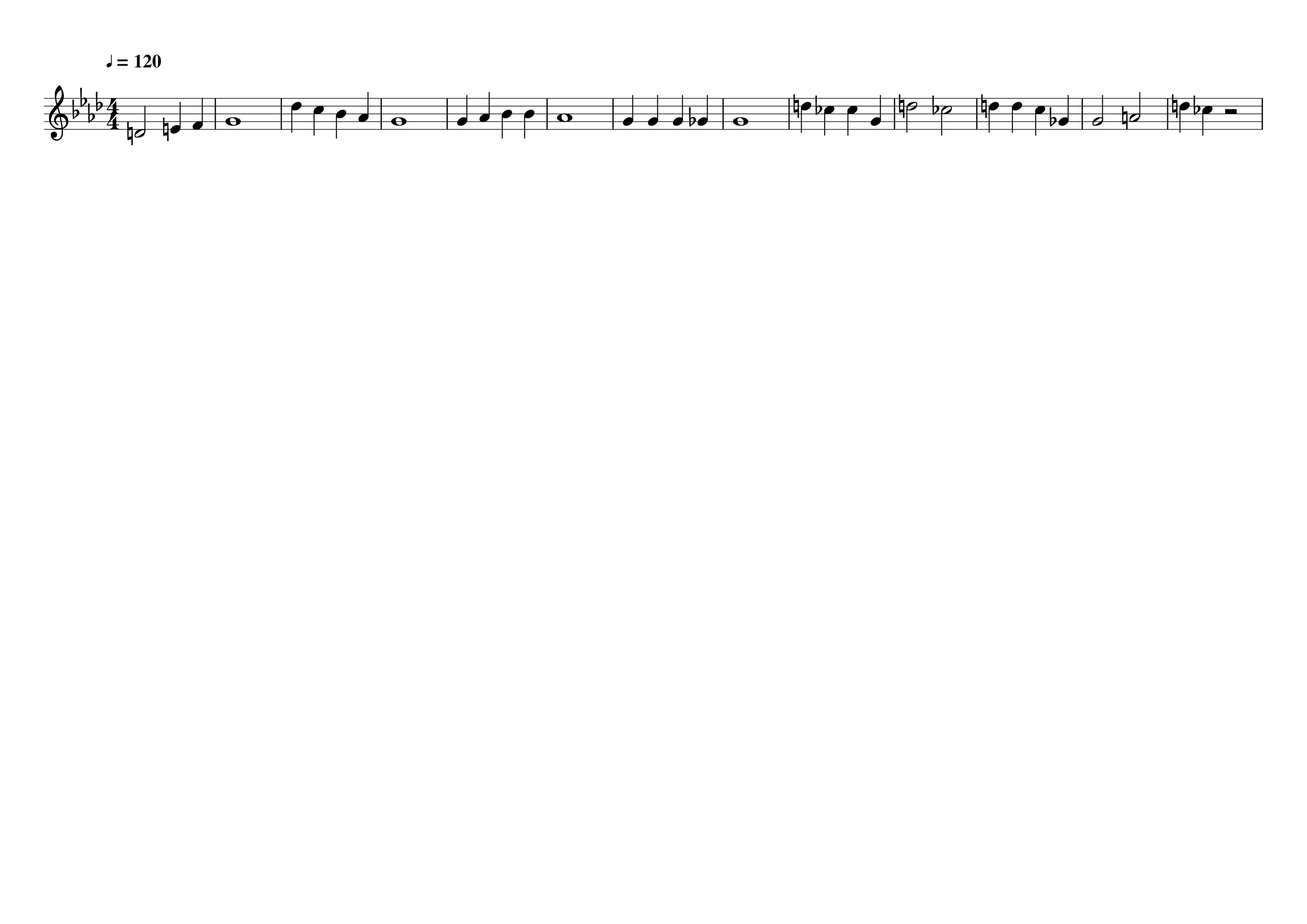}
        \caption{LSTM Model}
        \label{fig:db12lstmsong}
    \end{subfigure}
    
    ~ %add desired spacing between images, e. g. ~, \quad, \qquad, \hfill etc. 
    \begin{subfigure}[b]{\textwidth}
        \includegraphics[trim= 0 2000 0 200,clip,width=\textwidth]{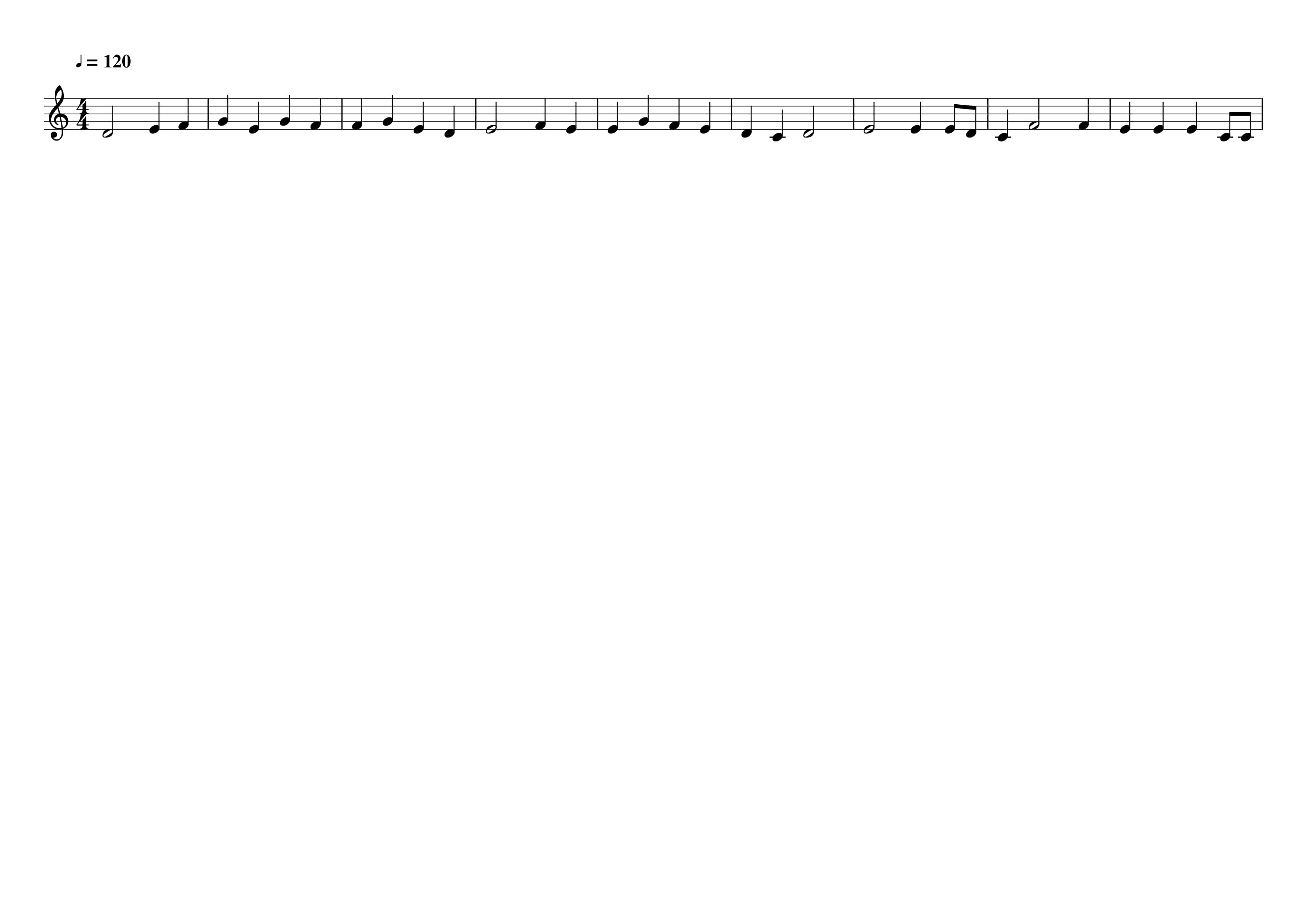}
        \caption{GRU Model}
        \label{fig:db12grusong}
    \end{subfigure} \\
    \caption{Generated Melodies DB12 Dataset}\label{fig:generatedmelodiesdb12}
\end{figure}

\begin{figure}[h!]
    \centering    
    \begin{subfigure}[b]{\textwidth}
        \includegraphics[trim= 0 2000 0 200,clip,width=\textwidth]{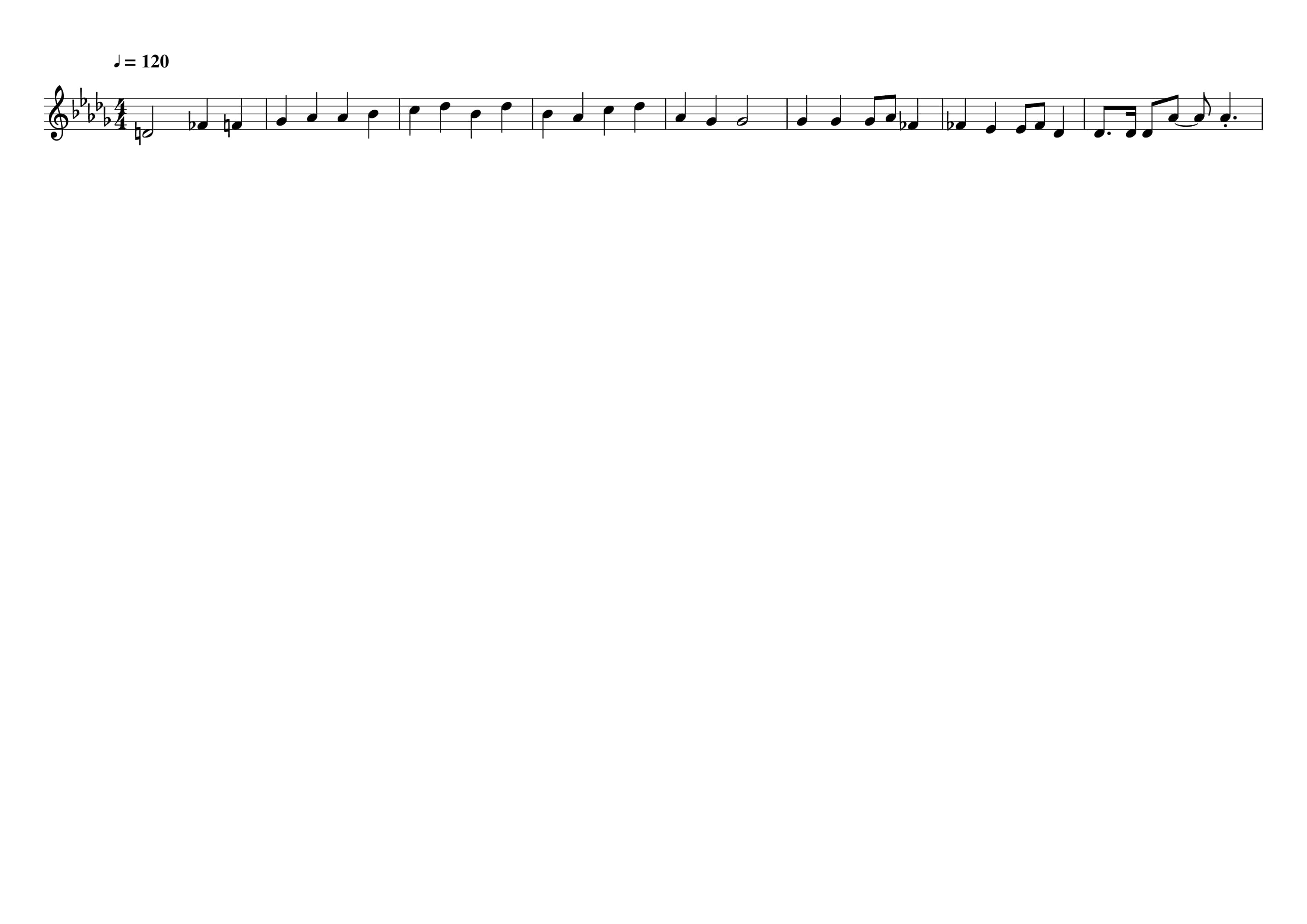}
        \caption{LSTM Model}
        \label{fig:intervallstmsong}
    \end{subfigure}
    
    ~ %add desired spacing between images, e. g. ~, \quad, \qquad, \hfill etc. 
    \begin{subfigure}[b]{\textwidth}
        \includegraphics[trim= 0 2000 0 200,clip,width=\textwidth]{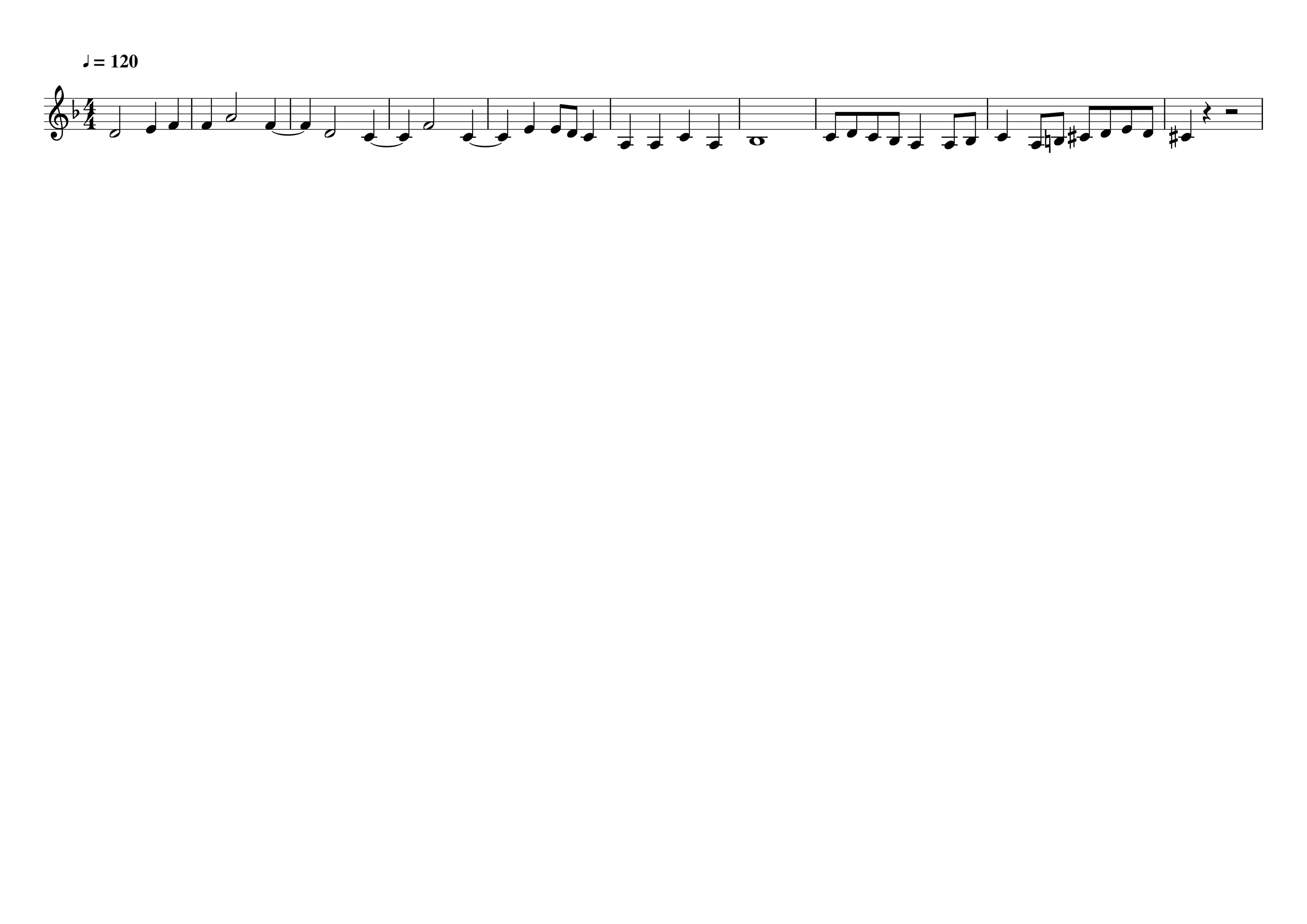}
        \caption{GRU Model}
        \label{fig:intervalgrusong}
    \end{subfigure} \\
    \caption{Generated Melodies Interval Dataset}\label{fig:generatedmelodiesinterval}
\end{figure}

Figures \ref{fig:generatedmelodiescontrol}, \ref{fig:generatedmelodiesdb12} and \ref{fig:generatedmelodiesinterval} shows the musical sheet of the 6 melodies from table \ref{tab:metricsMostRepresSongs}. Both CONTROL model songs have only natural notes and an evident high use of quarter notes, in the GRU case (Fig. \ref{fig:basegrusong}) the LM has relatively good results with a good balance of total pitches. 

The DB12-GRU model  (Fig. \ref{fig:db12grusong}) also shows no alterations and especially a good CMM with smooth changes across the whole piece and more variation in the rhythm using some half and eighth notes, however, the LM is the second worse between the models, in this case, because almost all the spans are below the minimum number of pitches (5), having as result a song with few pitches.

The DB12-LSTM (Fig. \ref{fig:db12lstmsong}) has, in this case, the higher centricity, the G4 quarter note repeated 3 times in bar 7 is a good example of why. This model also has good use of rhythms and phrase structure, you can describe the melody with an individual motif of 4 quarter notes followed by a whole note, and just some variations of this, the problem in this song is the distance in some changes like the jumps from a G4 to a D5 in bar 2, 8 and 9. This explains the score for the CMM, although this value is close to the expected one in the dataset evaluation (table \ref{tab:metricDataset}).

In the case of the INTERVAL models, as expected by the CMM and LM metrics (Tables \ref{tab:optimalnumberoflayers} and \ref{tab:metricsSelectedModel} ), The INTERVAL-LSTM (Fig. \ref{fig:intervallstmsong}) song has no abrupt changes from note to note and the range of pitches is well balanced. The INTERVAL-GRU song (Fig. \ref{fig:intervalgrusong}) has almost no quarter notes, and show some rhythmic and Melodic patterns even when this model didn't highlight in the metrics, in this model we can see also some half notes crossing bars in bars 2, 3 and 4.

\section{Conclusions and future research} \label{sec:conlusions}

About the 2 cell types, it is interesting that the representative melody of the LSTM model, when trained with the DB12 dataset, show a good learning capacity of phrase patterns. The GRU models are in general less stable than the LSTM ones (Sec \nameref{sec:experiments}), especially with the densest models. LSTM models have a more expected behavior, with a clear correspondence between the network depth and learning convergence, it is important to say though, that sections "\nameref{subsec:generatedmelodies}" and "\nameref{subsec:automEval100songs}" suggest that GRU models tend to explore more rhythmic patterns in the generation.

As found in the previous research, the depth of the network is not a trivial problem of using so many units as possible, section "\nameref{sec:experiments}" shows many examples where models with 512 and 1024 units have better results for the cost function than the ones with 2048.

Finally, concerning the dataset variations, the CONTROL dataset seems to have less exploratory models since they use only natural notes and a lot of quarter notes in the representative melodies. The DB12 dataset has the highest scores for the CENTR metric and in general shows more interesting musical patterns than the other representation alternative as intervals.

As future work. It would be interesting to see the performance of our optimal six models when used to generate sequences in other areas. Also, The implementation of the automatic metrics is a good starting point to make them more robust. These metrics could be used to train models using reinforcement learning.

\section{Acknowledgement}

Special thanks to the Center of Excellence and Appropriation on Big Data and Data Analytics (Alianza CAOBA).

\section{Supplementary Materials}
As support for the paper, there is a web page in the URL: \url{https://sebasgverde.github.io/rnn-time-music-paper/} with all the material necessary for research replication, which includes:

\begin{enumerate}
    \item Datasets
    \item Network Weights
    \item Midi songs of the final models
    \item RNN model
    \item Library for music evaluation
    \item Detailed instructions to run the scripts
\end{enumerate}

Scripts available include:
\begin{enumerate}    
    \item Environment setting
    \item Datasets and weights downloading
    \item Number of units experiment
    \item Generating songs with trained models
    \item Music Evaluation of models, songs and datasets
\end{enumerate}

\parskip 18pt

%References
\bibliographystyle{cmj}
\bibliography{bibliography}
\end{document}